# Symmetry-breaking Actuation Mechanism for Soft Robotics and Active Metamaterials

Shuai Wu[a,1], Qiji Ze[a,1], Rundong Zhang[a], Nan Hu[b], Yang Cheng[c], Fengyuan Yang[c], Ruike Zhao[a,*]

[a]Soft Intelligent Materials Laboratory, Department of Mechanical and Aerospace Engineering, The Ohio State University, Columbus, OH, 43210, USA;

[b]Department of Civil, Environmental and Geodetic Engineering, The Ohio State University, Columbus, OH, 43210, USA;

[c]Department of Physics, The Ohio State University, Columbus, OH, 43210, USA

[1]S. W. and Q. Z. contributed equally to this work.
[*]Correspondence and requests for materials should be addressed to R. Z. (zhao.2885@osu.edu)

**ABSTRACT**

Magnetic-responsive composites that consist of soft matrix embedded with hard-magnetic particles have recently been demonstrated as robust soft active materials for fast-transforming actuation. However, the deformation of the functional components commonly attains only a single actuation mode under external stimuli, which limits their capability of achieving tunable properties. To greatly enhance the versatility of soft active materials, we exploit a new class of programmable magnetic-responsive composites incorporated with a multifunctional joint design that allows asymmetric multimodal actuation under an external stimulation. We demonstrate that the proposed asymmetric multimodal actuation enables a plethora of novel applications ranging from the basic 1D/2D active structures with asymmetric shape-shifting to biomimetic crawling and swimming robots with efficient dynamic performance as well as 2D metamaterials with tunable properties. This new asymmetric multimodal actuation mechanism will open new avenues for the design of next-generation multifunctional soft robots, biomedical devices, and acoustic metamaterials.

# INTRODUCTION

Recent rapid advances in stimuli-responsive materials has enabled new devices and structures with programmable properties and configurations at a wide range of length scales[1-5]. Among a variety of stimuli-responsive materials[6, 7], magnetically actuated soft materials have recently shown diverse programmable shape transformations with promising applications, such as actuators[8], metamaterials[9], soft robots[10-12], and biomedical devices[13, 14]. **Figure 1a** shows a hard-magnetic soft material composed of a soft elastomer matrix with embedded hard-magnetic particles. When being magnetized under a large external magnetization field (~1.5 T), the material forms magnetic domains with strong remnant magnetization. Thereafter, under a small magnetic field (typically <200 mT), these domains align their magnetization with the applied magnetic field and induce instantaneous micro-torques that lead to a large mechanical deformation. In **Figure 1a**, the magnetic particles are magnetized in the horizontal direction. When a downward magnetic field is applied (**Figure 1b**), the material bends down. The deformation can be predicted via finite-element (FE) simulations based on the recent theoretical work on the mechanical behavior of hard-magnetic soft materials[15].

Under a switching magnetic field (from downward to upward), the actuation shows mirror symmetry. This has been used for interesting applications, such as providing propulsion force for soft robots[16-21]. However, the mirror symmetric deformation (or single modal deformation) has some limitations. In real world, it is highly desirable to achieve asymmetric responses under a symmetrically switching stimulus, as this can create a biased responsive loop. For example, in cyclic motions of bird flying or frog swimming, the asymmetric motions of wings or legs are the key to generate a net propelling, which can never be achieved through symmetric shape-shifting. The symmetry-breaking mechanisms could potentially bring new opportunities in designing multifunctional materials and structures. However, this is very challenging to achieve in smart materials, largely due to the material homogeneity and isotropy.

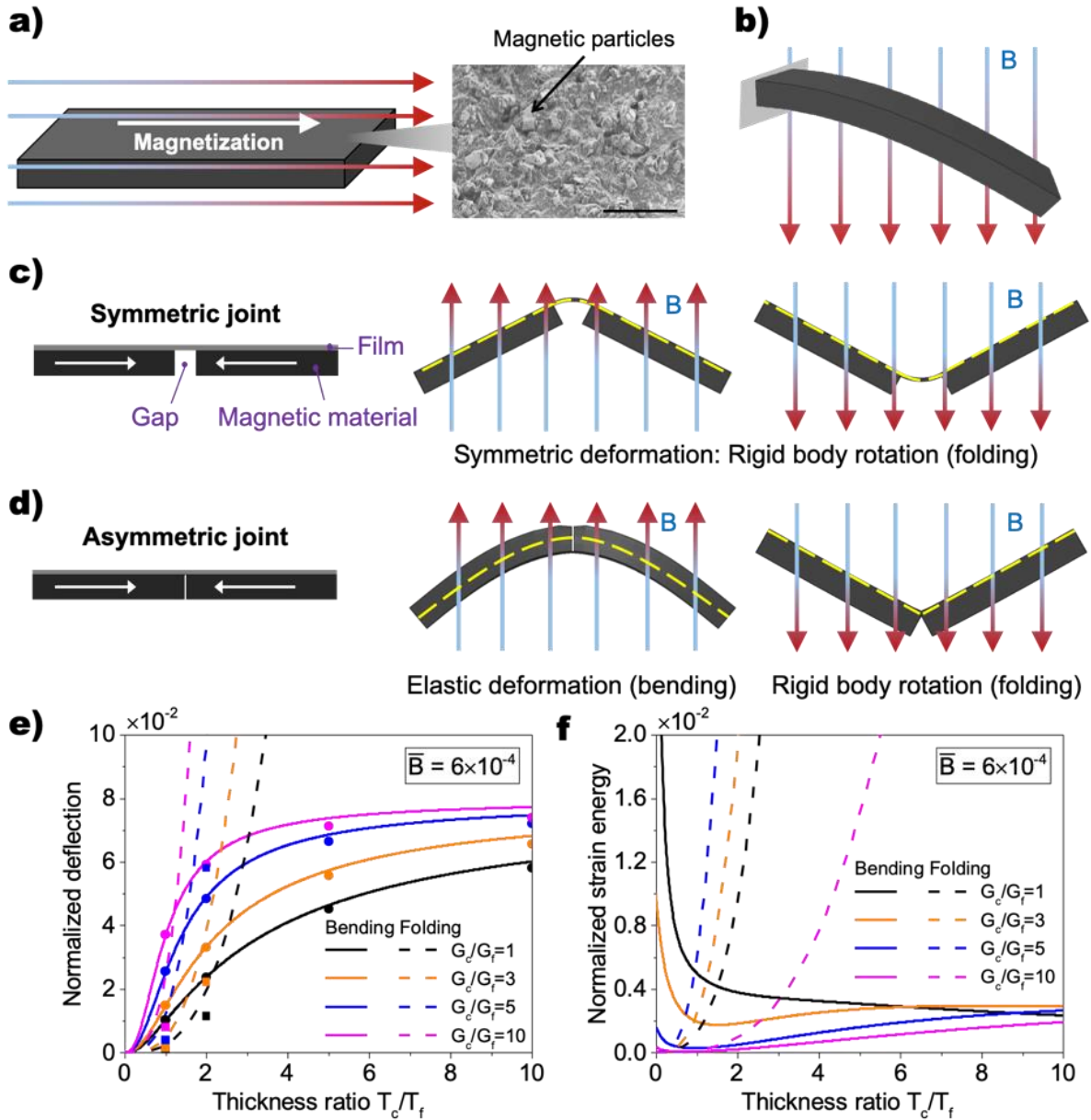

**Figure 1.** The hard-magnetic soft material and the multimodal deformation mechanism. (*a*) Schematics of a hard-magnetic soft material. Scale bar in the middle image, 30 μm. (*b*) Actuation of the hard-magnetic soft material. (*c*) Schematics of a symmetric joint. (*d*) Schematics of an asymmetric joint. (*e-f*) Theoretical calculations of (*e*) the normalized deflection and (*f*) the normalized strain energy with respect to the thickness ratio between the magnetic cell and the film.

To address this limitation, we propose a transformative mechanism for multimodal deformation that breaks the deformation symmetry through a mechanics-guided design of *shifting* *the neutral axis* in the material systems. Here, a *symmetry-breaking concept* is adopted to our new material system, which consists of two magnetic unit cells bonded by a soft thin film as either a symmetric joint (**Figure 1c**) or an asymmetric joint (**Figure 1d**). For the *symmetric joint* (**Figure 1c**), a gap is used between the two unit cells to avoid interference during the actuation. Under an external magnetic field, the neutral axis of the system (dashed yellow lines) is always along the thin film's mid-plane. A folding-like behavior is achieved through the rigid-body rotation of the magnetic unit cells and the elastic deformation of the film. Under a switching magnetic field, the deformation exhibits mirror symmetry. For the *asymmetric joint* (**Figure 1d**), the two adjacent unit cells are in direct contact. The neutral axis of the system is still along the film's centroid when the unit cells fold towards the film direction, similar to that in **Figure 1c**. However, upon switching the magnetic field, the neutral axis shifts to the mid-plane of the magnetic unit cells (see **Figure S3** for details). In this case, bending is determined mainly by the elastic deformation of the magnetic unit cells. With the asymmetric joint, we achieve multimodal deformation, i.e., *bending* (elastic deformation) and *folding* (rigid body rotation), in one material system by *shifting the neutral axis* between the mid-planes of the magnetic unit cells and the thin film. As demonstrated below, this new asymmetric joint approach enables an effective way of inducing *multimodal deformation* and leads to a wide-open design space for structures with tunable physical properties for a plethora of promising applications.

**RESULTS AND DISCUSSION**

**Theoretical analysis of the asymmetric deformation.** To investigate the behaviors of the asymmetric joint induced by a magnetic field applied perpendicular to the beam's programmed magnetization direction, we analytically calculate the deflections and strain energies (See *Supplementary Information* (SI) for details). Here, we define $G_f$, $G_c$, $T_f$, $T_c$ to be the shear moduli and thicknesses of the film and the magnetic unit cell, $L$, $W$, $M_c$ to be the length, width, and the magnetization of the magnetic unit cell, and $B$ to be the applied magnetic field. By normalizing the deflection and the applied magnetic field as $\overline{\upsilon^{\max}} = \upsilon^{\max}/L$ and $\overline{B} = M_c B/G_c$ and defining the modulus ratio and thickness ratio as $\alpha = G_c/G_f$ and $\beta = T_c/T_f$, the normalized deflections for bending and folding are

$$\overline{\upsilon^{\max}_{bend}} = \frac{400\alpha\beta^2\left(\alpha\beta^2 + \beta\right)}{3\left[\alpha^2\beta^4 + 1 + 4\alpha\left(\beta^3 + \beta + \frac{3}{2}\beta^2\right)\right]}\overline{B}; \quad (1)$$

$$\overline{\upsilon^{\max}_{fold}} = 4\alpha\beta^3 \overline{B}. \quad (2)$$

**Figure 1e** shows the normalized deflections as functions of the thickness ratio for bending (solid lines) and folding (dashed lines), respectively, under $\overline{B} = 6\times10^{-4}$. As the thickness ratio $T_c/T_f$ increases, both bending and folding deflections increase, but the bending deflection reaches a plateau. This is because of the significantly increased bending stiffness of the magnetic unit cells. In general, when $T_c/T_f > 2$, folding exhibits larger deflection than bending, indicating a strong asymmetric deformation under the switching magnetic field with the same magnitude. To validate the theoretical calculations, FE simulations (see **Figure S3** for details) are conducted and the results of bending (circles) and folding (squares) are also shown in **Figure 1e**, indicating good agreements. In addition, the strain energy stored in the system are calculated and normalized by $\overline{U} = U/U_{\mathrm{mag}}$ as plotted in **Figure 1f**, where $U$ is the total strain energy and $U_{\mathrm{mag}} = M_c B W T_c L$ is the magnetic potential defined as the work required to align magnetization with the applied magnetic field. For the folding behavior, the strain energy is mainly stored in the film and increases with the increasing thickness ratio due to the larger deflection. Based on the theoretical results, to achieve large symmetry-breaking deformations, we design the systems with the same magnetic unit cells ($G_c$ = 164 kPa, $L$ = 15 mm, $W$ = 7 mm, $T_c$ = 2mm and $M_c$ = 109 kA/m) and the soft film ( $G_f$ = 19 kPa and $T_f$ = 0.3 mm), leading to the modulus ratio of 8.6 and the thickness ratio of 6.7.

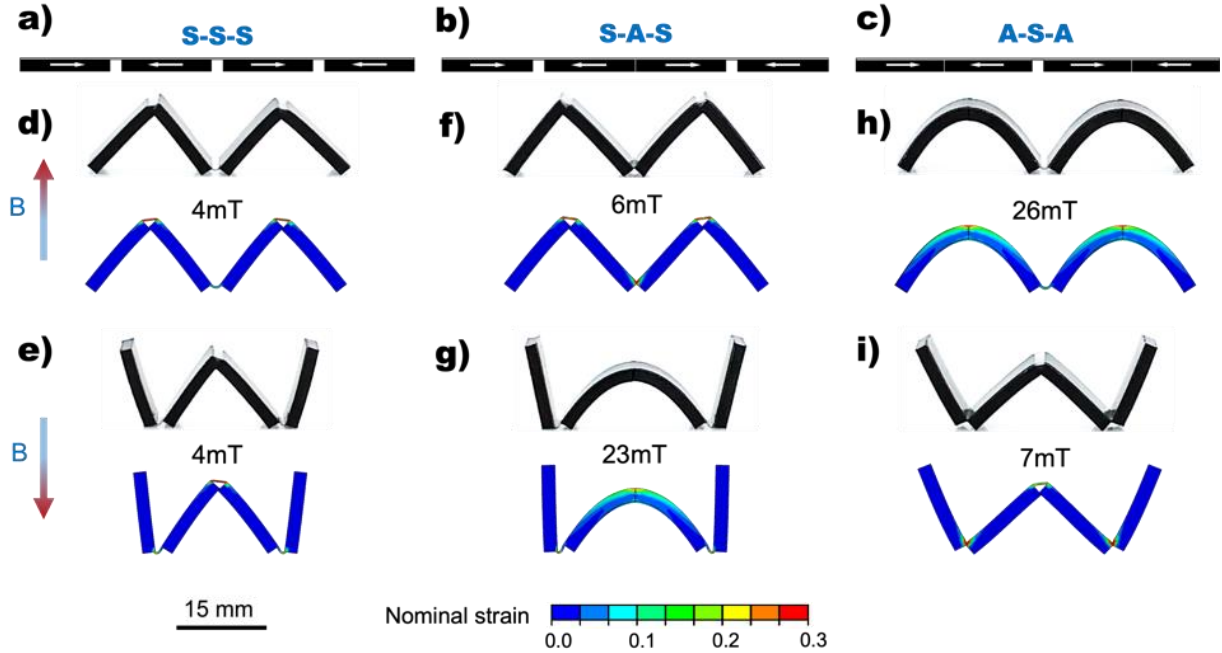

**Figure 2.** The magnetic-actuated multimodal deformations of straight four-cell systems with different combinations of symmetric (S) and asymmetric (A) joints. The upward magnetic field is defined as positive. (*a*) S-S-S type. (*b*) S-A-S type. (*c*) A-S-A type. (*d-e, f-g, h-i*) The experimental and simulated deformations for the design in (*a, b, c*).

**One-dimensional (1D) multimodal actuation.** To demonstrate the basic asymmetric deformation modes, the systems using three joints with a combination of symmetric (S) and asymmetric (A) ones are designed and fabricated as shown in **Figure 2** (see Materials and Methods for fabrication details): S-S-S, S-A-S, and A-S-A. The unit cells are magnetized in the horizontal direction with the same alternating magnetizations for all three joint combinations (**Figure. 2a-c**). A vertical uniform magnetic field $B$ is applied with upward as positive and downward as negative (see **Figure S4** for the electromagnetic coils as the source of the uniform magnetic field). The deformations of these systems under the applied magnetic field are predicted by FE simulations and experimentally implemented (see **Video S1**).

For the S-S-S combination (**Figure 2a**), either an "M" shape or a "W" shape is achieved through folding under the switching magnetic fields with amplitude $|B| = 4$ mT, giving a single deformation mode with a near mirror symmetry (**Figures. 2d & e**). The S-A-S combination (**Figure 2b**) uses one asymmetric joint in the middle, thus two deformation modes, an "M" mode under $B = 6$ mT (**Figure 2f**) and an arc mode under $B = -23$ mT (**Figure 2g**), are induced with the same deformation amplitude, which is defined as the vertical distance between peak and valley in the transformed shape. As the bending behavior indicates a larger bending stiffness than the folding, a larger magnetic field is required for bending to reach the same deformation amplitude (**Figure S3** for calculation of the asymmetric bending stiffness). Next, for the A-S-A combination with two asymmetric joints by the sides, we obtain a two-arc deformation mode (**Figure 2h**) under $B = 26$ mT, and a "W" mode with the same deformation amplitude under a much smaller magnetic field $B = -7$ mT (**Figure 2i**). **Figure 2d-i** and **Video S1** show the actuation.

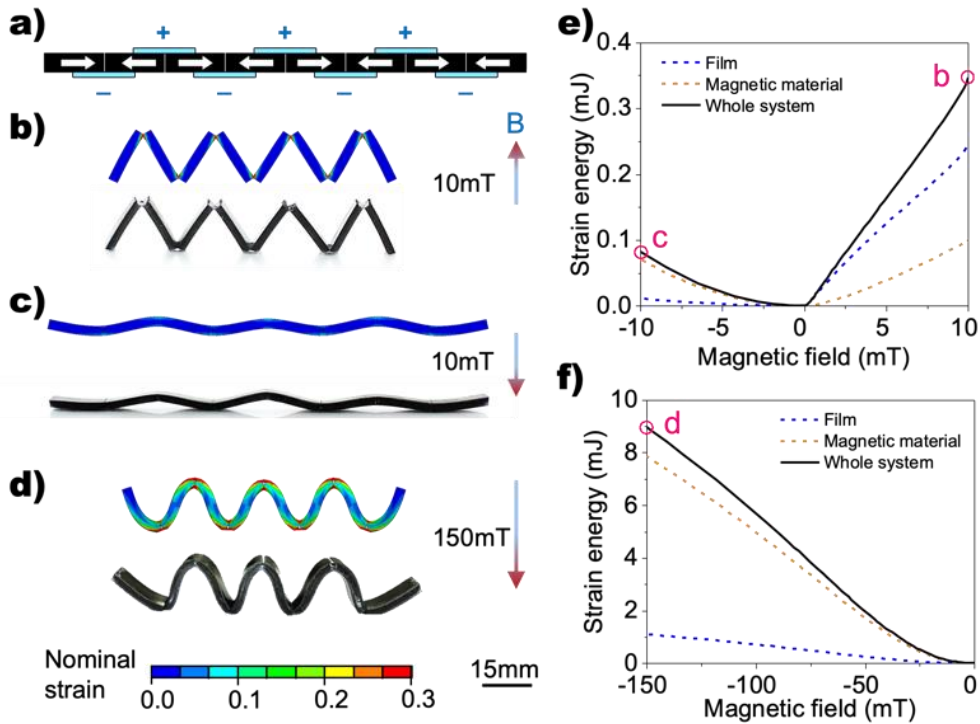

**Figure 3.** The magnetic-actuated multimodal deformations of a straight eight-cell system. (*a*) The schematic geometry and the programmed magnetizations. (b-d) Simulation and experimental validations of the eight-cell system under the applied magnetic field of (*b*) 10 mT, (*c*) -10 mT, and (*d*)-150 mT. (*e-f*) Strain energy stored in the film, the unit cells, and the whole system with respect to the applied magnetic field. Circles b, c, d represent the total strain energy of the deformed system in (*b*), (*c*), and (*d*) respectively.

**1D multimodal deformation by asymmetric joint allocation**. We now demonstrate that the asymmetric joint concept can be further exemplified by strategically tailoring the allocation of multiple joints. In the previous systems, all the magnetic unit cells are positioned on the same side to the film. Here, we alternate the position of the asymmetric joint to be either above (+) or below (-) the unit cells. **Figure 3a** illustrates a "-/+" type of joint allocation for a straight eight-cell system with alternating magnetizations in the horizontal direction. The deformation modes of the eight-cell system under different magnetic fields, $B$ = 10 mT, -10 mT, and -150 mT, are predicted by the FE simulations and validated by the experiments (**Figures 3b-d**; also see **Video S2**). When a positive magnetic field of 10 mT is applied, the joints fold, forming a zigzag shape. In addition, **Figure 3e** shows the quantitative study of the strain energy stored in the film (dashed blue curve), the magnetic cells (dashed orange curve) and the whole system (solid black curve), indicating the total strain energy of 0.35 mJ (point b), with most of the energy stored in the film.

When the magnetic field is reversed to $B$ = -10 mT (**Figure 3c**), the system transforms into a sinusoidal mode by bending. Under the same magnitude of the actuation field ($|B|$ = 10 mT), however, the deformation amplitude is only ~1/3 of that achieved by folding, due to the much larger bending stiffness of the magnetic cells. Moreover, it indicates that the strain energy is mostly stored in the magnetic cells (point c in **Figure 3e**), with a total strain energy of 0.08 mJ. Next, in order to obtain the same deformation amplitude by bending as that by folding under $B$ =10 mT, a much larger magnetic field, $B$ = -150 mT, is required to overcome the high bending stiffness, as shown in **Figure 3d**. In this case, the total strain energy (9.0 mJ) in the system (point d in **Figure**

**3f**) is 26 times of that in the folding with same amplitude (0.35 mJ), resulting in a much stiffer system. The insight of energy storage in the multimodal deformation suggests the potentials of developing new novel strategies for designing material systems with tunable physical properties while retaining the same overall geometrical deformation.

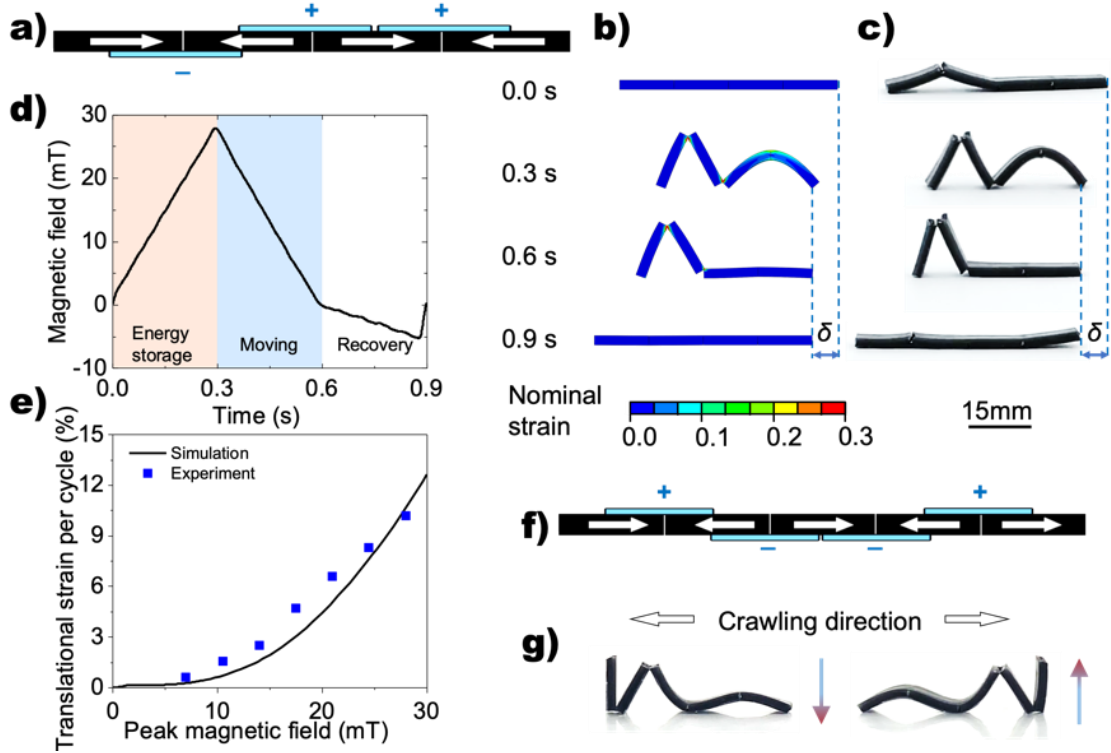

**Figure 4.** The design of biomimetic crawling robots with multimodal deformation. (*a*) The schematic geometry and programmed magnetizations. (*b*) The simulation of deformation for crawling motion with contour showing the nominal strain. (*c*) The experimental crawling motion. (*d*) The profile of the vertically applied magnetic field for motion control. (*e*) The translational strain in terms of the applied peak magnetic field. (*f*) The schematic geometry and programmed magnetizations for a straight dual-direction crawling robot. (*g*) The dual-direction crawling motion under the switching magnetic field.

**1D multimodal deformation for crawling motion.** The previous 1D configurations break symmetry in the actuation direction, but they still retain the symmetry with respect to the centroid (or self-symmetry). Such a self-symmetry leads to the contraction or extension with respect to the centroid, without creating an overall translational motion of the system. As discussed previously, breaking symmetry is highly desirable to generate a net gain in many nature systems. Here, we demonstrate that by breaking the self-symmetry and taking advantage of the stored elastic energy under magnetic field, the multimodal deformation can be utilized for translational motions, such as crawling. As shown in **Figure 4a**, a straight four-cell system is designed with the horizontally alternating magnetizations and an asymmetric joint allocation of "-/+/+". The crawling mechanism under a vertical magnetic field is illustrated in both simulation (**Figure 4b**) and experiment (**Figure 4c**; also see **Video S3**). As illustrated in **Figure 4d**, one crawling cycle consists of three steps: energy storage, moving, and recovery. When the magnetic field decreases to zero in the moving step, the elastically deformed magnetic unit cells release the stored elastic energy in the bending mode and recover to the flat state first. These flattened magnetic cells provide a large surface area in contact with the ground, and the resultant high friction prevents the system from moving backward when unfolding the left unit cells during the recovery step, thus creating a net forward motion. It should be noted that a small negative magnetic field is applied at the end to force the system to return to its initial flat configuration.

The synergic effort of the folding and bending behaviors, coupled with the change of contact conditions with the substrate, generates synchronous actuations that allow the crawling motion. The translational strain per cycle, which is defined as the crawling distance $\delta$ divided by the total length of the system (60 mm), is measured to be 10%, under the peak magnetic field of 28 mT. **Figure 4e** shows an increase of the translational strain with the increase of the peak magnetic field. The measured (blue squares) and the simulated (black curve) single-cycle translational strains are in good agreement. Guided by **Figure 4e**, by controlling the magnetic field's magnitude and frequency, we can realize a stable crawling robot with a speed up to 34 mm/s (0.56 body length/s), which is much faster compared with the reported crawling robots in the literatures (< 0.24 body length/s)[22-25].

Furthermore, the asymmetric actuation can be used to achieve a dual-direction crawling robot. Here, a five-cell system (with a total length of 75mm) with a "+/-/-/+" asymmetric joint allocation is designed as illustrated in **Figure 4f** (also see **Video S3**). Driven by a vertical magnetic field with a peak value of 30 mT, the dual-direction crawling robot can achieve a 13% translational strain for each full motion cycle. This new crawling mechanism by multimodal deformation reveals huge advantages including reconfigurable motion, large translational strain, and simple yet robust control for soft robots.

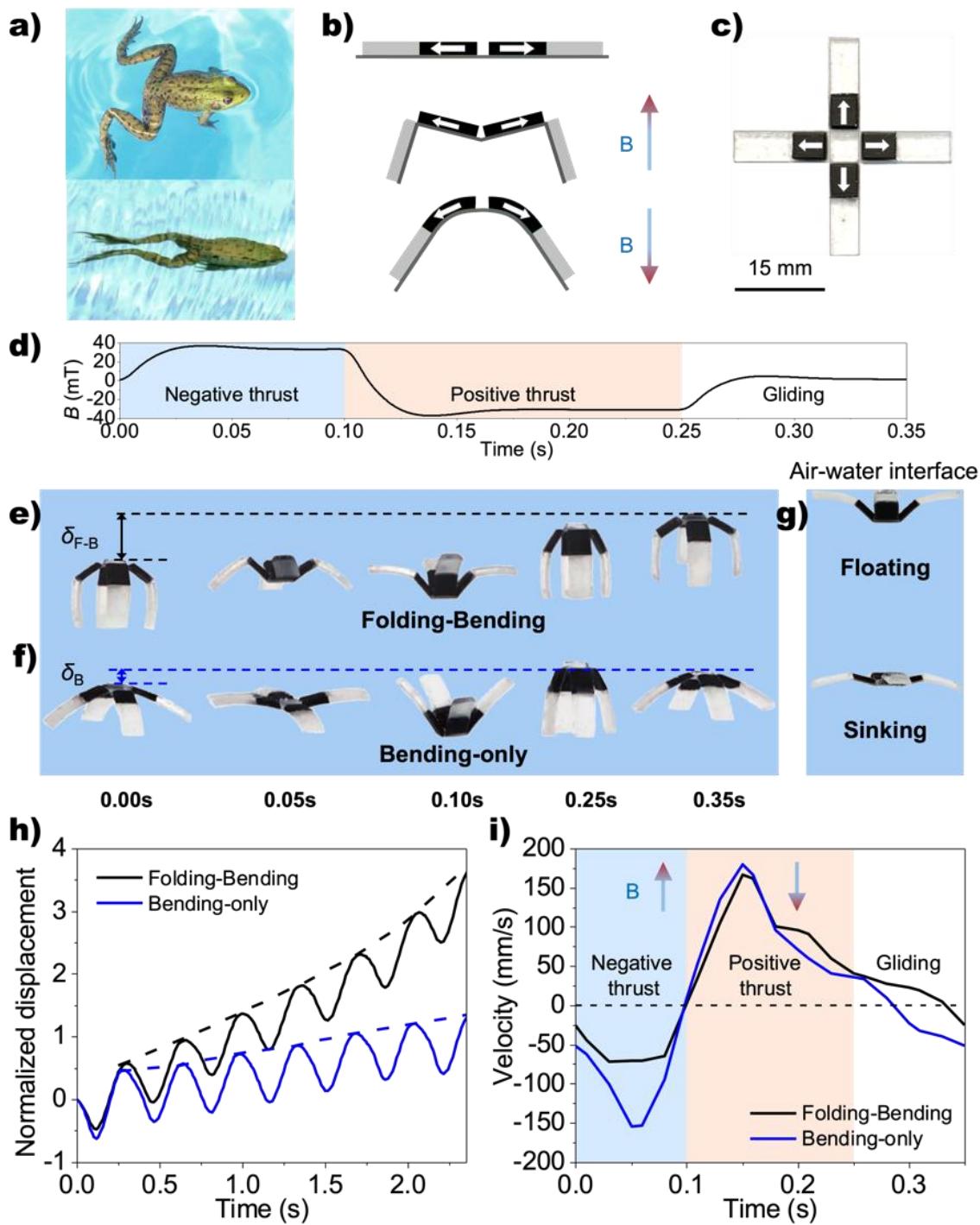

**Figure 5.** The design and characterization of a biomimetic swimming robot. (*a*) The propulsive mechanism of a swimming frog. (*b*) The schematic geometry and the programmed magnetizations of a four-cell system. (*c*) The fabricated four-leg swimming robot. (*d*) The control profile of the magnetic field. (*e-f*) The propulsive performance under one cycle of applied magnetic field for (*e*) the biomimetic folding-bending system with asymmetric joints and (*f*) the bending-only system. (*g*) The controlled floating and sinking motions. (*h-i*) A quantitative comparison between the biomimetic folding-bending system and the bending-only system in terms of (*h*) the normalized displacement under cyclic magnetic actuations and (*i*) the velocity profile in one actuation cycle.

**Asymmetric actuation for effective propulsion.** As discussed above, the ability of asymmetric shape-shifting can be easily found in nature and has brought new opportunities in designing functional devices[26-28] and robots[29-33]. Inspiring natural examples of multimodal deformation in locomotion can be commonly observed in animals. Taking frogs as an example, they generate thrust by pushing the water backwards using their hind limbs and webbed feet. For each propelling cycle, there are three phases: a folding phase with a negative thrust, an extending phase with a positive thrust, and a gliding phase that maximizes the positive thrust from the extending phase[34, 35]. Frogs can impressively achieve an effective net propelling (positive thrust - negative thrust) through the asymmetric thrusts[36], as shown in **Figure 5a**[37, 38]. Utilizing our asymmetric multimodal deformation, this strategy can be adopted to design a biomimetic swimming robot with effective net propulsion.

**Figure 5b** illustrates the biomimetic swimming legs with an A-S-A joint combination and the programmed magnetizations. The symmetric joint connects two magnetic unit cells working as the muscles and two non-magnetic cells on the sides. When the magnetic field is in the positive direction, the joints are in folding mode, which has a relatively low water resistance due to the low bending stiffness provided by the film and the small area exposed along the moving direction. When the magnetic field is switched to the negative direction, the joints switch to the bending mode, which ensures a large area as well as a stiff bending mode that pushes the water with a large reaction force for effective propelling. By such a simple design, our biomimetic swimming robot achieves asymmetric water resistance and thus a net propulsion during a motion cycle. To ensure stability in the water, we use a "four-leg" design with eight unit cells (two cells for each leg), as shown in the **Figure 5c** (see Materials and Methods for fabrication process). The applied magnetic field profile is illustrated in **Figure 5d**, including a negative thrust, a positive thrust and a gliding phase. In order to demonstrate the effective propulsion by the biomimetic multimodal deformation, we compare the performance of our biomimetic design with a control case that uses symmetric joints in the middle of each leg, as shown in **Figures 5e & f** and **Videos S4 & S5**.

To provide a quantitative comparison, we calculate the normalized displacement, defined as the actual displacement divided by the robot length (the maximum vertical extension, 15 mm), of both systems by measuring the robots' positions during multiple swimming cycles from **Video S5**. As shown in **Figure 5h**, after seven swimming cycles in a duration of 2.45s, the biomimetic system (black curve) reaches an effective displacement that is 2.5 times of the control system's displacement (blue curve). Additionally and more importantly, the net displacement (dashed curves) shows that the biomimetic system has an accelerated upward motion (the black dashed curve with a convex shape) whilst the control system shows a linear increase in the displacement (the straight blue dashed curve). We calculate the swimming velocities in a single swimming cycle for both cases. A significant difference in velocity between the biomimetic system (-70 mm/s) and the control system (-155 mm/s) in the negative thrust phase is observed (**Figure 5i**), indicating that the folding motion in the biomimetic robot can effectively reduce the negative thrust just as the frog does. Moreover, the biomimetic swimming robot can better take advantage of the positive thrust and glides at a higher velocity in the gliding phase, which further surpasses the control system.

In addition to swimming, more interesting motions like floating on the surface of the water and controlled sinking can be achieved, as shown in **Figure 5g** and **Videos S4 & S5**. To float on the water surface, the robot legs slightly fold up under a constant upward magnetic field (40 mT) to obtain a large contact area with the water surface and take advantage of the surface tension. By reducing the magnetic field (to 20 mT for example), the decreased surface tension can no longer hold the robot's body weight, thus a controlled sinking motion can be realized with desired velocity.

By utilizing multimodal deformation, our untethered biomimetic swimming robot provides an effective propulsion mechanism and outperforms many existing swimming robots. Notably, our swimming robot reaches a large average velocity of 23 mm/s and 1.53 times body length/s, which is much faster compared with the existing untethered stimuli-responsive swimming robots ($< 0.69$ body length/s)[39-41]. It should also be noted that our biomimetic swimming robot has an excellent propelling performance as it has to overcome its weight (the body has a density of 1.17 g/cm$^3$), where most reported swimming robots swim horizontally. Paired with the rational design of the multimodal deformation and the advanced control in 2D and 3D magnetic fields, we envision potential routes for developing biomimetic robots with complex yet controllable motions and the ability for completing multitasks at different length scales.

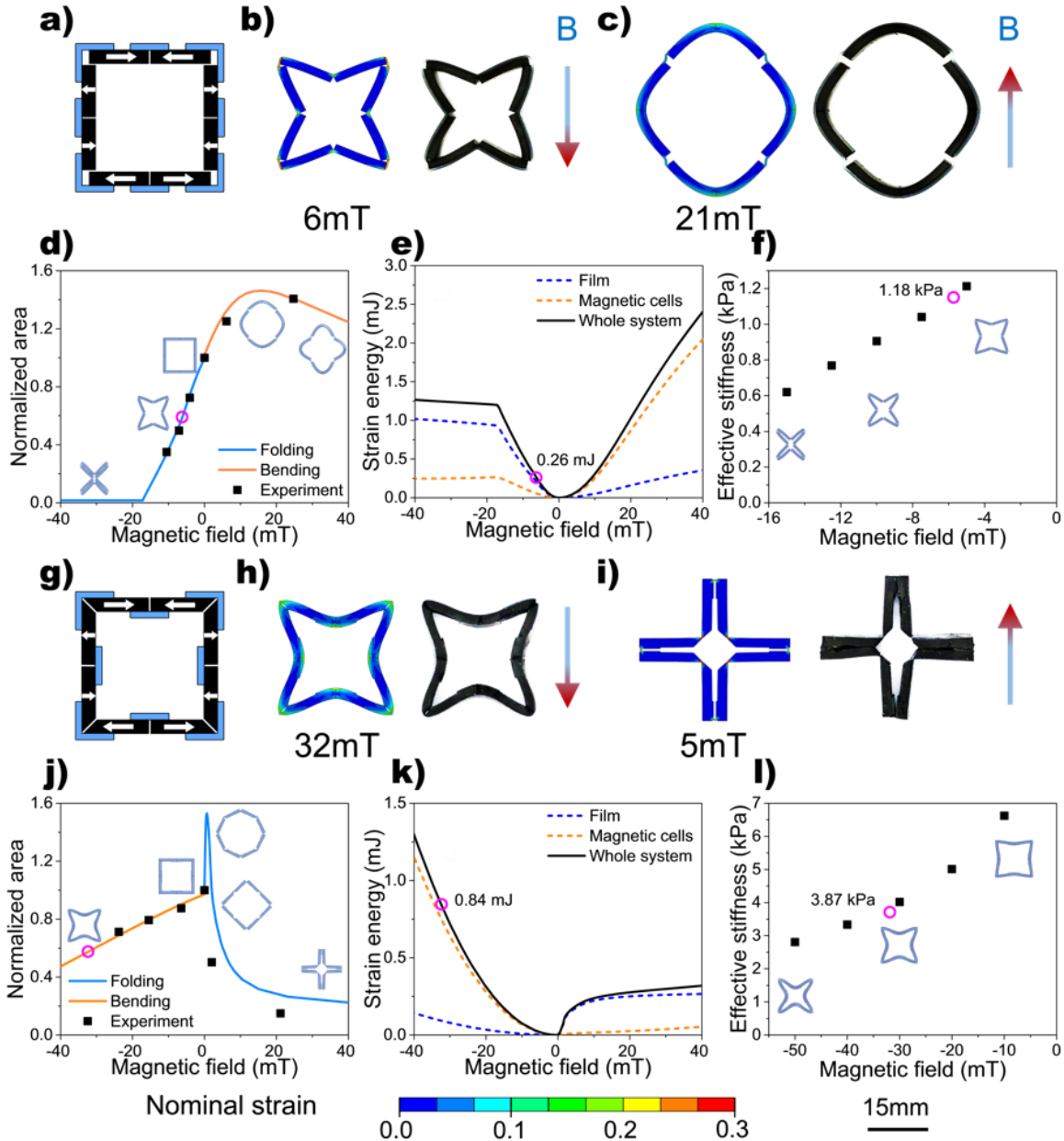

**Figure 6.** Magnetic-actuated multimodal deformations for 2D systems. (*a-f*) A square system with outside symmetric and asymmetric joints: (*a*) The schematic geometry and programmed magnetizations; (*b-c*) the simulations and experimental validations of the deformations; (*d*) the normalized inner area; (*e*) the strain energy; and (*f*) the effective stiffness. (*g-l*) A square system with alternating inside and outside joints: (*g*) the schematic geometry and programmed magnetizations; (*h-i*) the simulations and experimental validations of the deformations; (*j*) the normalized inner area; (*k*) the strain energy; and (*l*) the effective stiffness.

**Two-dimensional (2D) multimodal system**. We now explore two 2D geometries with different joint combinations and allocations that lead to shape transformations and tailorable properties. We start with a square ring with the joint allocation and magnetizations illustrated in **Figure 6a**. The adjacent magnetic cells are programmed to have opposite magnetizations in the horizontal direction, connected by the joints on the outside. Under a negative magnetic field, the system contracts to a star shape by folding inward (**Figure 6b**). When a positive magnetic field is applied, the system bends outward to form a rounded shape with a continuous curvature (**Figure 6c**). In **Figure 6d**, the normalized area, defined as the enclosed area at the deformed configuration divided by the initial one, is calculated from the FE simulation. Varying the magnetic field from -40 mT to 40 mT, the original square transforms to a close-packed "X" shape as the extreme state in the negative-field regime, and to a "flower" shape in the positive-field regime (see **Video S6**). The normalized area decreases linearly with the negative magnetic field and reaches its minimum (almost zero) at $B = -17$ mT where the magnetic cells are in contact. When applying an increasing positive magnetic field, the normalized area increases to its maximum at $B = 16$ mT, which is 1.5 times of the initial area, and then decreases with the increasing magnetic field. Five experimental measurements of the inner area are shown by the black squares in the **Figure 6d**, with excellent agreements with the simulation. **Figure 6e** shows the calculation of the total strain energy during the multimodal actuation. The effective stiffness of the deformed system also varies with the applied magnetic field, as shown in **Figure 6f** (see *SI* for the details on calculation of the effective stiffness). Within a small range of the applied magnetic field from -5mT to -15mT, the effective stiffness of the system decreases by 49% (from 1.21 kPa to 0.62 kPa).

Next, a different set of multimodal deformation is demonstrated by using a square ring with same overall dimensions and magnetizations, but a different joint allocation (**Figure 6g** and **Video**

S6). In **Figure 6g**, four asymmetric joints are located on the inner surfaces of the square ring. Under a negative magnetic field (**Figure 6h**), the ring transforms into a similar star shape by the elastic deformation, instead of folding shown in **Figure 6b**. By switching the field direction, the magnetic cells fold to an octagon first and eventually turn to a close-packed cross shape at $B = 5$ mT (**Figure 6i**). The normalized area and shape configurations during the actuation are shown in **Figure 6j**. The largest area (1.5 times of the initial area) is obtained when the system deforms to an octagon shape under a very small magnetic field $B = 0.6$ mT. This configuration is not stable and instantaneously transforms to the close-packed cross shape, due to the magnetic attraction between the adjacent unit cells. The smallest area is then reached around $B = 5$ mT at the close-packed state, revealing only 20% of the initial area, and the strain energy in the system reaches a plateau (**Figure 6k**). For the effective stiffness (**Figure 6l**), it shows the similar trend, i.e. as the system shrinks under the negative magnetic field, the effective stiffness decreases.

For the same "star" shape, the bending one (**Figure 6b**) attains a higher total strain energy (0.84 mJ) while the folding one (**Figure 6h**) has a lower total strain energy (0.26 mJ), as indicated in **Figures 6k&e**, respectively. In addition, the effective stiffness by bending (3.87 kPa) is 3.3 times of that by folding (1.18 kPa). These results indicate that the same geometries achieved by different multimodal designs could have different stored energy and different stiffnesses, which can be used in the rational design of novel magnetic-actuated metamaterial system with unprecedented tunable properties, which will be demonstrated below.

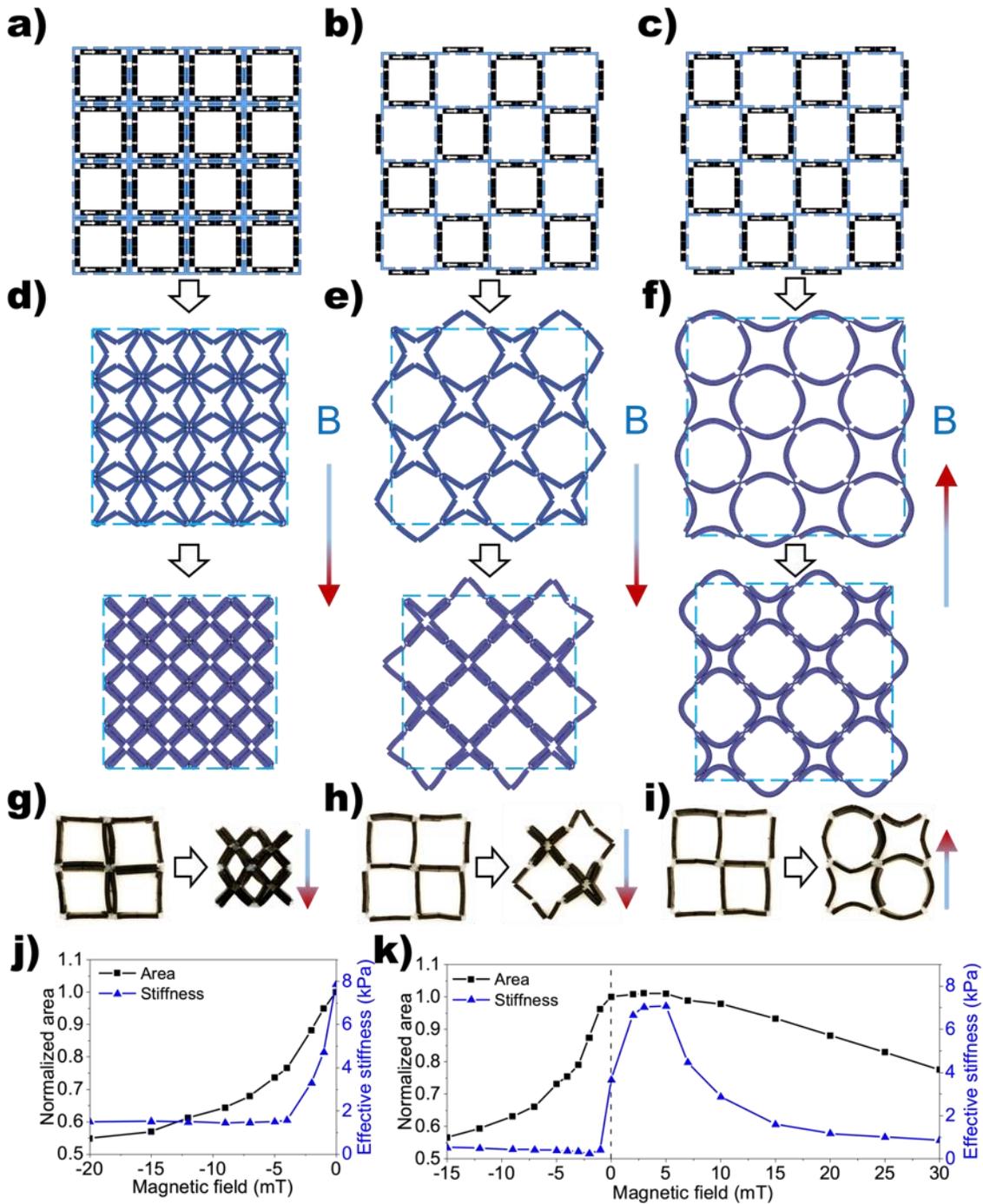

**Figure 7.** Magnetic-actuated metamaterial systems. (*a-c*) Schematic geometries and programmed magnetizations of the metamaterial systems. (*d-f*) Simulations of the deformations of the four-by-four metamaterial systems under magnetic stimulations. (*g-i*) Experimental validations on the deformations of two-by-two systems. (*j-k*) The normalized area (black) and the effective stiffness (blue) under controlled magnetic field.

**Active metamaterial with tailorable properties**. To prove the concept of using the multimodal system for new active metamaterial systems with tunable properties, we assemble the square unit cells to four-by-four metamaterial systems and explore their property change under magnetic stimulations. As shown in **Figure 7a**, the square design in **Figure 6a** is stacked together to form a four-by-four metamaterial system. When a negative magnetic field is applied, all square unit cells contract at the same time in both the horizontal and vertical directions, behaving as an auxetic material with a negative Poisson's ratio (**Figure 7d**, and **Video S7**)[42]. As we further increase the magnetic field, each square unit cell forms a close-packed cross shape and the entire structure reaches its minimum area, 54% of the initial area, which is close to the theoretical calculation of 50%. A two-by-two system is also fabricated and actuated under the same magnetic field used in the simulation, as shown in **Figure 7g**, confirming the simulation results. The blue dashed-line box shown in **Figures 7d-f** connects the four vertices of the four-by-four metamaterial system, and the area inside is used to characterize the system's total area during deformation. Quantitative studies from the FE simulations on the normalized area and the effective stiffness are plotted in **Figure 7j**. As the applied magnetic field increases, the area of the whole structure decreases monotonically, while the stiffness drops significantly at the beginning and becomes stable when the magnetic field is larger than 5 mT.

As seen above, when we place square unit cells next to each other, the adjacent cells directly contact. This prevents the expansion of the whole system. To overcome this limit, another active metamaterial system is introduced by placing the same square unit cell at the interval of its own dimension (**Figures 7b & c**). This allows both folding (**Figure 7e**) and bending (**Figure 7f**) under switching magnetic fields (see **Video S7**). Again, the experimental results of the two-by-two structure confirm the simulation predictions (**Figures 7h & i**). The calculations on the normalized area and the effective stiffness under the applied magnetic fields are plotted in **Figure 7k**. Under either a negative or positive magnetic field, the area of the whole system varies in the range of 55%-100% of the initial area. The effective stiffness decreases significantly under the negative field when folding is induced. However, under the positive magnetic field, the effective stiffness first increases significantly and then decreases as buckling starts to occur. Overall, under the magnetic field, the stiffness of the metamaterial system can be actively tuned from 7% to 200% of its initial value, which is a much wider controllable range as compared with the reported metamaterial systems [43-45]. Although we only show the results of two simple designs, we envision that the introduced active metamaterial system opens avenues to a new class of material system with many other interesting multifunctionality such as controllable acoustic properties and energy storage capabilities.

**CONCLUSION**

In summary, the multimodal deformation of the magnetic-responsive soft materials is demonstrated through the introduction of the asymmetric joint, which breaks the intrinsic symmetry of the material system by shifting the neutral axis under directional actuations. The multimodal deformation is further harnessed by the design of soft robots with effective motion and demonstration of active metamaterials with tunable properties. While in this paper we focus on simple geometrical forms, the proposed concept of multimodal deformation can be transformed into a variety of complex 3D geometries by strategically assembling the basic structures. Empowered by the emerging 3D printing technology[46-52], it becomes highly feasible to fabricate complex multi-material architectures at a wide range of length scales. In addition, the mechanics-guided design approach is a pressing necessity as we enter this new era with increasing demands on reconfigurable materials and systems[53, 54]. The growing body of research implies a new movement of integrated design methodology, where properties can be tailored individually at the local/mechanism level and then achieve an enhanced performance or unprecedented functionalities at the global/system level[55, 56]. We envision that this transformative approach will result in a variety of routes in designing programmable machines and systems in the near future.

**Materials and Methods**

*Fabrication of the basic material system:* The basic material system consists of magnetic unit cells and soft thin films made from silicone rubbers. The magnetic unit cells were fabricated by mixing PDMS (Sylgard 184, DowCorning, Inc., Midland MI, USA, weight ratio of PDMS base and curing agent as 20:1) with NdFeB microparticles (Magnequench International, LLC) with a volume fraction of 20%. The mixture was then poured into a mold and cured in the oven at 70°C for 30 minutes. The thin film was fabricated by spin-coating a layer of Ecoflex (Ecoflex$^{TM}$ 00-30, Smooth-On, Inc., Macungie, PA, USA) at 400 rpm for 20 s on top of the magnetic materials. The resultant thickness of the thin film was measured to be 0.3 mm. The whole system was cured again in the oven for 30 minutes at 70°C to ensure chemical crosslinking.

*Fabrication of biomimetic swimming robot:* The body of the biomimetic swimming robot was fabricated in a 3D-printed mold (Formlabs, Inc., Somerville MA, USA). The magnetic and non-magnetic unit cells were fabricated by using PDMS with a weight ratio of base and curing agent to be 15:1. For the magnetic unit cells, NdFeB powder with a volume fraction of 3% was added to the mixture before curing. A layer of Ecoflex thin film was spin-coated under the rate of 450 rpm for 30 s, giving a measured thickness of 0.15 mm. The density of the whole robot was measured to be 1.17 g ml$^{-1}$.

*Mechanical and magnetic characterization of materials:* The shear moduli of the fabricated material parts were measured by using a universal material testing machine (Model 3340, Instron, Inc, Norwood, MA, USA). The samples were uniaxially stretched at a low strain rate $0.01 s^{-1}$. The moduli were obtained by fitting the measured stress-stretch curves by the neo-Hookean model (see Figure S5, Supporting Information). The measured shear moduli of the magnetic PDMS and non-magnetic Ecoflex were 164 kPa and 19 kPa, respectively. The magnetization of the materials was measured by the vibrating sample magnetometer (VSM, 7707A Lake Shore Cryotronics, Inc., Westerville, OH, USA). The magnetic hysteresis loops were obtained, indicating the magnetization for the magnetic materials (PDMS 20:1 NdFeB 20 Vol %) to be 109 kA m$^{-1}$ (see Figure S6, Supporting Information).

*Finite element simulation:* To predict the magnetic-actuated deformation of the hard-magnetic soft active material, a user-defined 8-noded element subroutine (UEL) was implemented through the commercial software ABAQUS 2018 (Dassault Systèmes, France) [7, 8e]. A neo-Hookean energy function with coupled magnetic potential was used to describe the non-linear behavior of the material system. The UEL subroutine utilizes the materials' shear modulus, magnetization vector, and external field vector as the inputs to calculate the deformation of the material under magnetic actuation.

**ASSOCIATED CONTENT**

**Supporting Information**

The Supporting Information is available free of charge on the ACS Publications website at DOI: xxx.

Additional data, figures including: Analytical solution for bending and folding behavior of asymmetric and symmetric joints, calculation of the effective stiffness of the magnetic-actuated metamaterial, Simulations of a two-layer composite beam bending, Image of the single-axis

Helmholtz coils for generation of 1D magnetic field *B* with controllable magnitude and frequency, Mechanical characterization of materials used for demonstrations in the paper (PDF).

Video S1. Magnetic-actuated 1D multimodal deformation for the four-cell systems (mp4).

Video S2. Magnetic-actuated 1D multimodal deformation for the eight-cell system (mp4).

Video S3. Magnetic-actuated crawling robots (mp4).

Video S4. A four-leg biomimetic swimming robot with multimodal deformation (mp4).

Video S5. Comparison between the biomimetic swimming robot with multimodal deformation and the control case with bending-only deformation (mp4).

Video S6. Magnetic-actuated 2D multimodal deformation (mp4).

Video S7. Magnetic-actuated metamaterial systems (mp4).

**AUTHOR INFORMATION**

**Corresponding Author:**

Email: zhao.2885@osu.edu (R. Z.).

**Author Contributions**

S. W., Q. Z, R.Z. and R.Z. conceived the research. Q. Z., R. Z, Y. C. and F. Y. performed the experiments and analyzed the experimental results. S.W. and N.H. conducted finite element simulation, S.W. and R. Z. composed the manuscript. All authors reviewed the manuscript.

**Notes**

The authors declare no competing financial interest.

**Acknowledgements**

The research was supported by the Haythornthwaite Foundation Research Initiation Grant and OSU Institute for Materials Research. F.Y.Y. and Y.C. acknowledge the support from US Department of Energy under Grant No. DE-SC0001304.